\documentclass[twocolumn]{aastex61}
\usepackage{amsmath}
\usepackage{gensymb}
\accepted{June 15, 2018}
\submitjournal{ApJ}

\shorttitle{Optical and Infrared Photometry of the Nearby SN 2017cbv}
\shortauthors{Wee et al.}

\begin{document}

\title{Optical and Infrared Photometry of the Nearby SN 2017cbv}

\correspondingauthor{Jerrick Wee}
\email{wee.jerrick@u.yale-nus.edu.sg}

\author{Jerrick Wee}
\affil{Yale-NUS College, 
16 College Avenue West,
Singapore 138527, Singapore}

\author{Nilotpal Chakraborty}
\affiliation{Yale-NUS College, 
16 College Avenue West,
Singapore 138527, Singapore}

\author{Jiayun Wang}
\affiliation{Yale-NUS College, 
16 College Avenue West,
Singapore 138527, Singapore}

\author{Bryan Penprase}
\affiliation{Soka University of America, 
1 University Drive,
Aliso Viejo, CA 92656, USA}

\begin{abstract}

On 2017 March 11, the DLT40 Transient Discovery Survey discovered SN 2017cbv in NGC5643, a Type 2 Seyfert Galaxy in the Lupus Constellation. SN 2017cbv went on to become a bright Type Ia supernova, with a $V_{max}$ of 11.51 $\pm$ 0.05 mag. We present early time optical and infrared photometry of SN 2017cbv covering the rise and fall of over 68 days. We find that SN 2017cbv has a broad light curve $\Delta m_{15}(B)$ = 0.88 $\pm$ 0.07, a $B$-band maximum at 2457840.97 $\pm$ 0.43, a negligible host galaxy reddening where $E(B-V)_{host}$ $\approx$ 0, and a distance modulus of 30.49 $\pm$ 0.32 to the SN, corresponding to a distance of $12.58_{-1.71}^{+1.98}$ Mpc. We also present the results of two different numerical models we used for analysis in this paper: SALT2, an empirical model for Type Ia supernova optical light curves that accounts for variability components; and SNooPy, the CSP-II light-curve model that covers both optical and near-infrared wavelengths and is used for distance estimates. 

\end{abstract}



\section{Introduction} \label{sec:intro}

Type Ia supernovae (SNe) are hypothesized to be the consequence of a thermonuclear explosion of a carbon-oxygen white dwarf (WD) \citep{1982ApJ...253..798N, 1984ApJ...286..644N, 1986LNP...255...91W, 2000ARA&A..38..191H, 2008NewAR..52..381P} or a merger of a pair of white dwarfs in a close binary system \citep{1984ApJ...277..355W, 1984ApJS...54..335I}. In the former model, the WD accretes material from a non-degenerate secondary companion till it reaches close to the Chandrashekhar mass limit, after which carbon ignites, resulting in the explosion. Different models of this nature can be differentiated on the basis of the secondary companion. The white dwarf can accrete in the following ways: one, from a wind, the ``symbiotic channel" \citep{1992ApJ...397L..87M}; two, by mass transfer from a helium star, also known as the ``helium star channel" \citep{1982ApJ...253..798N, 2010A&A...523A...3L, 2003A&A...412L..53Y} or three, by the Roche-lobe Overflow, the ``RLOF channel" \citep{1992A&A...262...97V}. The merger of a double-degenerate pair is caused by either the shrinkage of their orbit due to gravitational radiation \citep{1984ApJS...54..335I, 1984ApJ...277..355W} or  perturbations from additional celestial bodies close to the binary system \citep{2013ApJ...766...64S, 2013MNRAS.435..943P, 2014MNRAS.439.1079A}.

SNe Ia are important objects in astronomy and cosmology as they serve as ``standard candles" \citep{1998AJ....116.1009R, 1999AIPC..478..129P}, which help us calculate vast astronomical distances. SNe Ia produce regular patterns in their light curves over the course of a few months, and these patterns have allowed SNe Ia researchers to determine the peak luminosity, host galaxy extinction, and correspondingly, the distances to these objects. The peak luminosity of SNe Ia can be determined by various methods, most notably through variations of the Philips relation \citep{1996AJ....112.2391H, 1999AJ....118.1766P, 2004A&A...415..863G, 2006ApJ...647..501P} and the multicolor light-curve shape (MLCS) method \citep{1996ApJ...473...88R, 1998AJ....116.1009R, 2007ApJ...659..122J}.

The color curves of SNe Ia are important in determining host galaxy extinction. While SNe Ia differ in the shape of their light curves $\Delta m_{15}(B)$ = 0.75 to 1.94 \citep{2003AJ....125..166K}, the color curves of all SNe Ia are thought to evolve in the same way 35 days after $T(B_{max})$, an epoch known as the Lira law regime \citep{1995}. The use of $B-V$ color curves is helpful in determining extinction as the color excess $E(B-V)$ of the SNe Ia can be derived with the data of Galactic dust reddening from the dust maps from \citet{1998ApJ...500..525S} and \citet{2011ApJ...737..103S}. However, extinction calculations with optical colors where $A_V = R_V \times E(B-V)$ translate to huge uncertainties in $A_V$ if $B-V$ has a substantial uncertainty. Extinction calculations would improve if both optical and near-infrared photometry are used \citep{2007AJ....133...58K}. 

The use of both optical and near-infrared (near-IR) photometry of SNe Ia has shown to be crucial in determining extinction in the line of sight of a Type Ia supernova \citep{2008ApJ...689..377W, 2009ApJ...704..629M, 1985ApJ...296..379E}. Particularly, the $V-K$ colors is especially useful for determining extinction as it is the most uniform of all color curves of SNe Ia \citep{1985ApJ...296..379E}.

With data on both the peak luminosity and extinction of a Type Ia supernova, we can derive distance modulus. The distance derived using the Type Ia supernova can also be used to cross-check with other cosmological standard candles in the galaxy, such as the TRGB method and Cepheid variables, which allows us to obtain a precise distance to the galaxy. While there are a lot of data on high-$z$ SNe ($z > 0.01$), data on low-$z$ SNe remain scarce, which introduces substantial errors to the precision of the Hubble constant in the local SNe \citep{2017arXiv170310616J}. Data of low-$z$ SNe's distances are critical in constraining the Hubble flow, and hence important to constrain the rate of expansion of the universe.

In this paper, we present early time optical and infrared photometry of the low-$z$ SN 2017cbv. SN 2017cbv (SN) was discovered by S. Valenti et al.\footnote{https://wis-tns.weizmann.ac.il/object/2017cbv/discovery-cert} on 2017 March 10 UT (= Julian Date 2,457,822.671), at right ascension 14\textsuperscript{h}32\textsuperscript{m}34\textsuperscript{s}.420 and declination of $-44^{\circ}$08'02 738". The host galaxy of the SN is a Type 2 Seyfert galaxy NGC5643. The galaxy has a redshift of 0.003999 and its heliocentric velocity is 1199 km s\textsuperscript{-1} \citep{2004AJ....128...16K}. Coincidentally, NGC5643 was also the host to another supernova, SN 2013aa, discovered in 2013 \citep{2014MNRAS.443.1849S}. Spectroscopic data of SN 2017cbv by Hosseinzadeh et al.\footnote{https://wis-tns.weizmann.ac.il/object/2017cbv/classification-cert} from the Las Cumbres Observatory determined SN 2017cbv to be a Type Ia supernova two weeks before maximum light.

\section{Observations} \label{sec:obs}

We used the Yale SMARTS 1.3m Telescope at the Cerro Tololo Inter-American Observatory (CTIO) for optical and near-infrared (near-IR) observation in $BVRIYJHK$ bandpasses. The images were taken with the ANDICAM instrument, an imager permanently mounted on the 1.3m telescope that takes simultaneous optical and infrared data. The ANDICAM instrument uses the standard Johnson $BV$ filters, Kron-Cousins $R$ and $I$ filters, standard CIT/CTIO $JHK$ filters, and a 1.05$\mu$m $Y$ filter. The optical field of view is 6.3' x 6.3', while the IR field of view is 2.34' x 2.34'. With the CCD readout in 2 x 2 binning mode, the ANDICAM instrument gives a plate scale on the 1.3m telescope of 0.369" pixel\textsuperscript{-1} for optical imaging and 0.274" pixel\textsuperscript{-1} for near-IR imaging. Further information can be found on the ANDICAM instrument specification website\footnote{http://www.astronomy.ohio-state.edu/ANDICAM/detectors.html}. 

\begin{figure}
\plotone{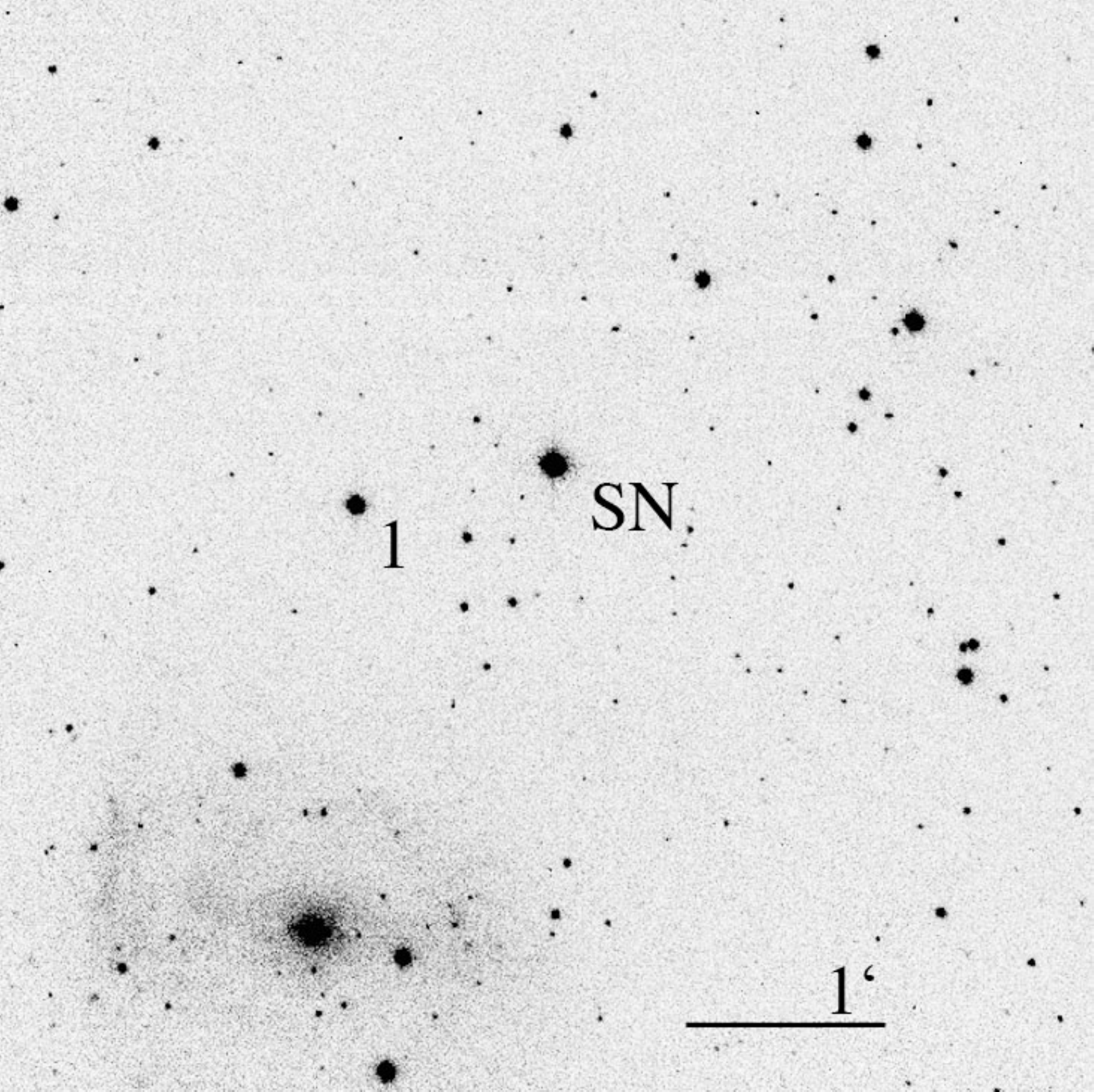}
\caption{V-band image of SN 2017cbv off NGC 5643 obtained with the CTIO 1.3m telescope on 2017 March 27. The exposure time was 15s. The local photometric standard is numbered. The bar corresponds to 1'. North is up, and east to the left. The supernova is marked 'SN', in the center of the image.\label{fig:imageop}}
\end{figure}

\begin{figure}
\plotone{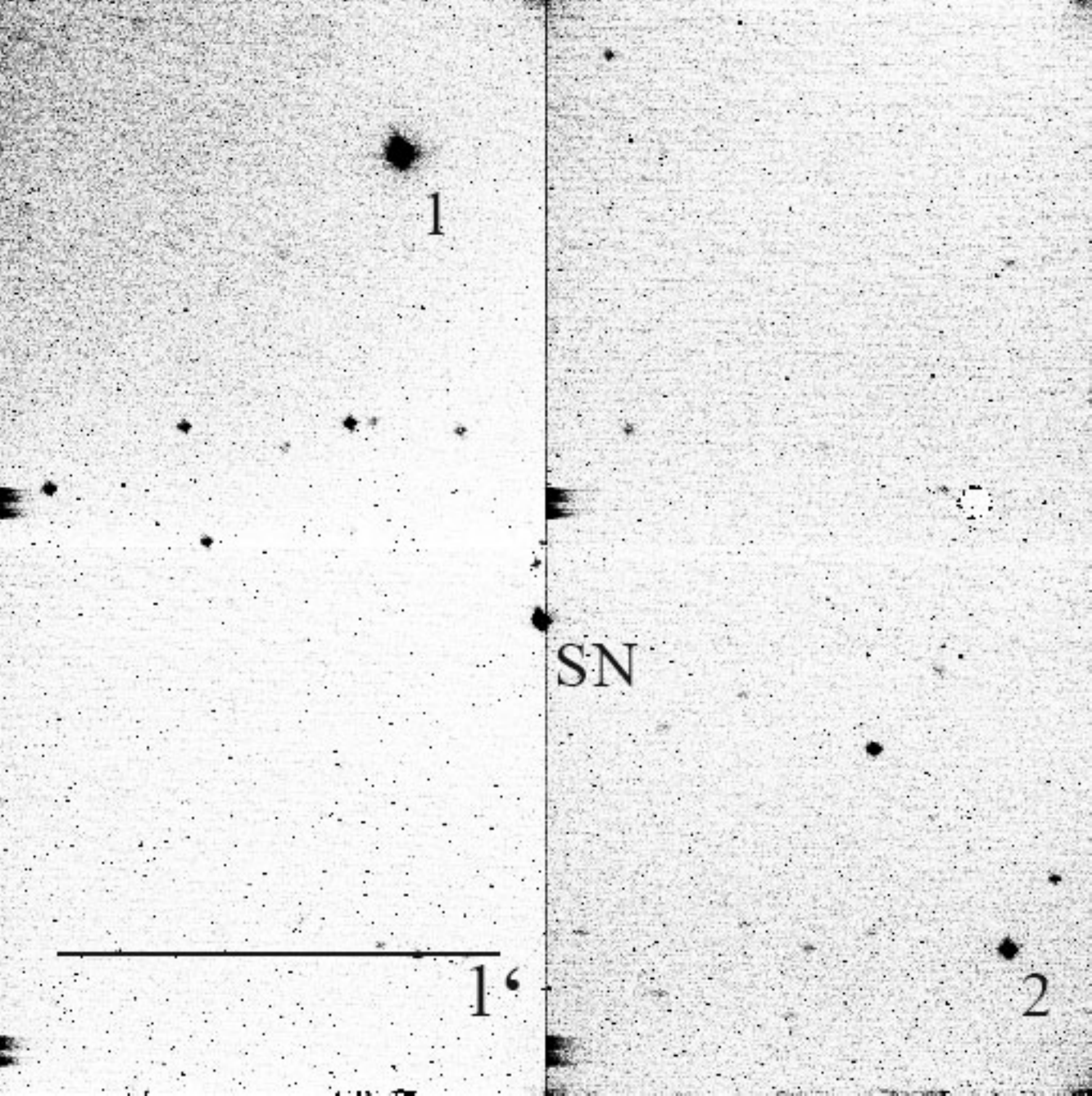}
\caption{J-band image of SN 2017cbv off NGC 5643 obtained with the CTIO 1.3m telescope on 2017 March 27. The exposure time was 15s. The local photometric standards are numbered. The bar corresponds to 1'. The supernova is marked 'SN', in the center of the image. North is right, east is down.\label{fig:imageir}}
\end{figure}

We started taking optical images of the SN on 2017 March 14, and started taking infrared images on 2017 March 18. Our observations ended on 2017 June 9, for a total of 68 nights of data. Figure \ref{fig:imageop} is a $V$-band optical image in which we identify the location of the SN and the local field standard, with Table \ref{tab:opseq} presenting the optical photometric sequence of the local field standard. Figure \ref{fig:imageir} is a $J$-band near-IR image which shows the location of the SN and the field standards. The data on the local field standards for the near-IR can be found on Table \ref{tab:irseq}.

We conducted the photometric reduction of the optical data using the Photutils's aperture photometry tool \citep{2016ascl.soft09011B}, a part of the Python-based package Astropy \citep{2013A&A...558A..33A}. The use of aperture photometry over point spread function (PSF) fitting is justified on the account that the SN is a low-$z$ supernova and that the SN appears at a substantial angular distance (2.57') from its host galaxy. This angular separation from the host galaxy reduces the chance that SN would be contaminated by the other sources of photons in the aperture radii. 

We estimated the magnitudes for a set of reference stars in the field of view of the SN using the mean value of five nights of observations. Our photometry for individual nights was computed using the grid of reference stars in the field, and these differentiated magnitudes were converted to $BVRI$ magnitudes using zero points derived from observations of photometric standards, described in Section 2.1. 

While there appears to be multiple bright stars in the field of view for the optical images, we have two problems with using multiple reference stars. One, our field of view is not consistent as we changed our coordinates throughout our observation. That severely limits our available reference stars to a small field of view centered on the galaxy. Two, depending on the exposure for a particular night, the reference stars available have a signal-to-noise ratio that is 5 to 8 times lower than the SN and Star 1, which reduces the quality of our light curves and increases errors in our results. For these reasons, we did not use more reference stars for our optical photometry than the ones labeled.

\begin{figure}
\plotone{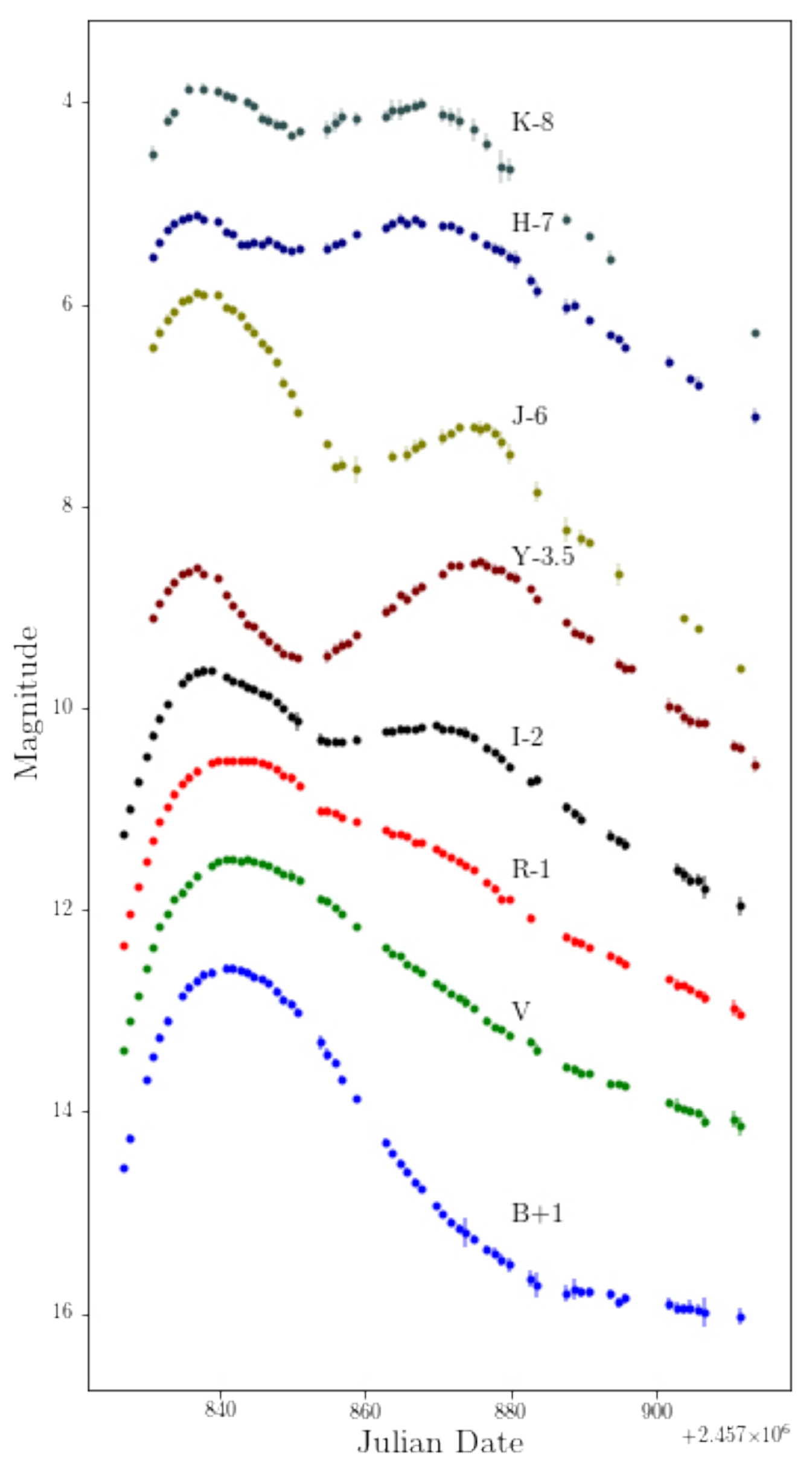}
\caption{Light curves of SN 2017cbv in the optical and near-infrared, with respective offset.\label{fig:fullc}}
\end{figure}

The near-IR data are reduced using the astronomy software Cyanogen Imaging MaxIm DL's photometric tool, which has the in-built capability to identify, track, and photometer objects across images using aperture photometry. We used Star 2 (Figure \ref{fig:imageir}) for $YJH$ filter magnitudes and Star 1 (Figure \ref{fig:imageir}) for $K$ filter magnitudes for our differential photometry. We present the $BVRIYJHK$ light curves in Figure \ref{fig:fullc}. 

We performed a Lomb-Scargle periodicity analysis on both reference stars with 152 and 149 photometric $V$-band data points for Star 1 and Star 2 respectively over 1798 days from the Catalina Sky Survey \citep{2009ApJ...696..870D}. Both stars have significant false alarm probabilities that makes them highly unlikely to be variables. We also conducted a $\chi^2$ fit test with a non-varying model for both reference stars and found that the $\chi^2$ fit was consistent with a model of non-variability. Both reference stars have been extensively observed by the Catalina Sky Survey and were not determined to be variables \citep{2017MNRAS.469.3688D}. The combination of our tests and the lack of variability found for both of our reference stars from \citet{2017MNRAS.469.3688D} gives us good confidence that the two reference stars are not variables.

Both random errors from our sky and background subtraction and systematic errors in zero points were computed in our systematic error budget. 

\begin{deluxetable*}{ccccccc}
\tablecaption{Optical Photometry Sequence Near SN 2017cbv \label{tab:opseq}}
\tablewidth{0pt}
\tablehead{
\colhead{Star ID} &
\colhead{$\alpha$ (J2000)} &
\colhead{$\delta$ (J2000)} & 
\colhead{$V$} & 
\colhead{$B-V$} &
\colhead{$V-R$} &
\colhead{$V-I$}
}
\startdata
  SN & 14\textsuperscript{h}32\textsuperscript{m}34\textsuperscript{s}.420 &  $-44^{\circ}$08'02 738" & ... & ... & ... &... \\
  1 & 14\textsuperscript{h}32\textsuperscript{m}39\textsuperscript{s}.982 & $-44^{\circ}$08'17 340" &12.806 (045)& 1.206 (113) & 0.7276 (085)&1.364 (103)\\
\enddata
\end{deluxetable*}

\begin{deluxetable*}{ccccc}
\tablecaption{Infrared Photometry Sequence Near SN 2017cbv \label{tab:irseq}}
\tablewidth{0pt}
\tablehead{
\colhead{Star ID} &
\colhead{$Y$} &
\colhead{$J$} & 
\colhead{$H$} & 
\colhead{$K$} 
}
\startdata
  1 & ...&...&...& 9.564 (064)\\
  2 & 12.993 (033)&12.706 (045)&12.495 (037)&  ...\\
\enddata
\end{deluxetable*}

\subsection{Zero Points}

We carried out standard calibrations for the optical zero points using the Landolt standards, namely: Ru149, Ru149A, Ru194B, and Ru194C \citep{1992AJ....104..340L}. Optical observations of the Landolt standard field were made with the same telescope and camera at CTIO over 5 nights between 2017 March 24 and 2017 April 21. The calibrations involved analyzing the intensities of the 4 standards stars each night alongside the airmass for calculating extinction.

From the dataset of each standard star, we arrived at an average zero point and its standard deviation. The zero points of the $BVRI$ bands were calibrated with extinction and color term corrections, given in Equation (1). In Table \ref{tab:exc}, we present the extinction and color term coefficients we used for our calibrations and the error of the data involved.

\begin{equation}
ZP_i = m_{inst_i} - M - E_i (X-1) - C_i (m_{inst_a} - m_{inst_b})
\end{equation}
where $ZP_i$ is the zero point of a filter $i$, $m_{inst_i}$ is the instrumental magnitude for a given filter $i$, $M$ is the literature magnitude, $E_i$ is the extinction coefficient for a given filter $i$, $C_i (m_{inst_a} - m_{inst_b})$ is the color term coefficient for a given filter $i$.

We obtained the extinction\footnote{http://www.ctio.noao.edu/noao/content/13-m-smarts-photometric-calibrations-bvri} and color term\footnote{http://www.ctio.noao.edu/noao/content/photometric-zero-points-color-terms} coefficients from the CTIO's calibration pages. The final zero points and the error used are the mean values of the zero points and standard deviations calculated from the independent observations of the different standard stars. 

The zero points of the $YJHK$ bands were calibrated using the Persson standard system \citep{1998AJ....116.2475P} with the P9144 field. The calibration involved independent observations of reference stars taken during 7 nights  between  2017 March 21 and 2017 June 9. The $Y$ magnitudes for the P9144 field in the Persson standard system are obtained from \citet{2017AJ....154..211K}. The procedure for data analysis and aperture choice of the near-IR zero points were the same as those of the optical zero points. 

In Table \ref{tab:zp}, we present the zero points of the $BVRIYJHK$ bands. The complete optical and near-IR photometric cadence is listed in Table \ref{tab:cadence}.

\begin{deluxetable}{ccc}
\tablecaption{Optical Extinction and Color Term Coefficients\label{tab:exc}}
\tablehead{
\colhead{Band} & 
\colhead{Extinction} & 
\colhead{Color Term}
}
\startdata
$B$ & 0.251 & $-0.079 (B-V)$ \\
$V$ & 0.149 & $0.018 (B-V)$ \\
$R$ & 0.098 & $-0.03 (V-R)$ \\
$I$ & 0.066 & $0.045 (V-I)$\\ 
$Y$ & 0.083  & -\\ 
$J$ & 0.010  & $-0.034(J-H)$\\ 
$H$ & 0.055  & $0.022(J-H)$\\ 
$K$ & 0.085  & $-0.003(J-K)$\\  
\enddata
\end{deluxetable}

\begin{deluxetable}{ccccc}
\tablecaption{Calibrated Zero Points of the $BVRIYJHK$ Bands\label{tab:zp}}
\tablehead{
\colhead{Filter} & 
\colhead{Zero Point} & 
\colhead{Error} &
}
\startdata
  $B$ &  22.019 & 0.054\\
 $V$ & 21.758 & 0.030\\
$R$ & 21.559 & 0.022\\
$I$ & 20.939 & 0.032\\
$Y$ & 18.274 & 0.022\\
$J$ & 18.177 & 0.028\\
$H$ & 18.305 & 0.030\\
$K$ & 17.764 & 0.037\\ 
\enddata
\end{deluxetable}

\begin{deluxetable*}{cccccccccc}[b!]
\renewcommand{\arraystretch}{0.75}
\tabletypesize{\footnotesize}
\tablecaption{Optical and Near-IR Photometric Cadence of SN 2017cbv \label{tab:cadence}}
\tablecolumns{10}
\tablehead{
\colhead{Julian Date} &
\colhead{Exposure} &
\colhead{} & 
\colhead{} &
\colhead{} & 
\colhead{} & 
\colhead{} &
\colhead{} &
\colhead{} & 
\colhead{} \\
\colhead{(2457000)} &
\colhead{Time ($V$)} &
\colhead{$B$} & 
\colhead{$V$} &
\colhead{$R$} & 
\colhead{$I$} & 
\colhead{$Y$} &
\colhead{$J$} &
\colhead{$H$} & 
\colhead{$K$} 
}
\startdata
826.759 & 15 & 13.557(064) & 13.398(041) & 13.352(033) & 13.255(054) & .... & .... & .... & ....\\
827.742 & 15 & 13.253(067) & 13.103(041) & 13.038(032) & 12.997(057) & .... & .... & .... & ....\\
828.782 & 30 & .... & 12.844(042) & 12.776(035) & 12.725(057) & .... & .... & .... & ....\\
829.781 & 30 & 12.685(067) & 12.572(046) & 12.518(033) & 12.483(054) & .... & .... & .... & ....\\
830.763 & 30 & 12.462(069) & 12.366(044) & 12.311(034) & 12.266(054) & 12.598(035) & 12.433(046) & 12.531(032) & 12.518(083)\\
831.773 & 30 & 12.263(063) & 12.172(048) & 12.128(049) & 12.102(065) & 12.470(030) & 12.277(049) & 12.380(037) & ....\\
832.755 & 30 & 12.095(065) & 12.035(045) & 11.985(038) & 11.961(063) & 12.339(038) & 12.153(051) & 12.269(037) & 12.176(064)\\
833.730 & 15 & .... & 11.902(052) & 11.845(047) & .... & 12.247(032) & 12.076(040) & 12.201(038) & 12.102(048)\\
834.677 & 15 & 11.857(069) & 11.824(056) & 11.758(051) & 11.762(079) & 12.175(033) & 11.965(041) & 12.153(044) & ....\\
835.783 & 15 & 11.766(079) & 11.749(057) & 11.681(053) & 11.695(079) & 12.144(034) & 11.947(045) & 12.139(039) & 11.860(052)\\
836.759 & 15 & 11.707(087) & 11.658(058) & 11.616(053) & 11.650(085) & 12.117(031) & 11.885(039) & 12.116(041) & ....\\
837.747 & 15 & 11.640(088) & .... & .... & 11.636(086) & 12.178(031) & 11.905(042) & 12.152(043) & 11.862(051)\\
838.802 & 15 & 11.616(088) & 11.563(054) & 11.539(048) & 11.635(074) & .... & .... & .... & ....\\
839.772 & 15 & .... & 11.512(047) & 11.530(039) & .... & 12.204(031) & 11.906(038) & 12.175(037) & 11.884(045)\\
840.768 & 15 & 11.575(085) & 11.509(047) & 11.516(039) & 11.698(060) & 12.373(033) & 12.024(044) & 12.272(035) & 11.940(044)\\
841.807 & 15 & 11.586(077) & 11.508(048) & 11.516(044) & 11.730(059) & 12.477(035) & 12.061(046) & 12.313(038) & 11.958(045)\\
842.789 & 15 & 11.594(076) & 11.511(047) & 11.522(039) & 11.757(060) & 12.562(036) & 12.121(048) & 12.406(037) & ....\\
843.791 & 15 & 11.614(082) & 11.505(047) & 11.512(038) & 11.791(059) & 12.661(030) & 12.222(044) & 12.398(034) & 11.999(047)\\
844.733 & 15 & 11.664(073) & 11.521(046) & 11.516(038) & 11.811(058) & 12.697(032) & 12.274(047) & 12.392(038) & 12.031(046)\\
845.756 & 15 & 11.691(074) & 11.540(048) & 11.541(039) & 11.862(057) & 12.765(034) & 12.384(043) & 12.400(042) & 12.155(048)\\
846.665 & 15 & 11.733(075) & 11.555(048) & 11.561(039) & 11.883(058) & 12.833(033) & 12.443(044) & 12.361(039) & 12.178(064)\\
847.755 & 15 & 11.806(078) & 11.609(049) & 11.612(040) & 11.949(057) & 12.895(033) & 12.582(043) & 12.408(044) & 12.216(057)\\
848.757 & 15 & 11.891(079) & 11.637(049) & 11.659(040) & 12.007(058) & 12.965(037) & 12.773(052) & 12.445(040) & 12.229(050)\\
849.735 & 15 & 11.940(081) & 11.662(116) & 11.698(118) & 12.080(058) & 12.976(030) & 12.873(039) & 12.464(040) & 12.328(046)\\
850.793 & 15 & 12.017(082) & 11.703(094) & 11.771(093) & 12.135(179) & 13.009(034) & 13.063(045) & 12.449(037) & 12.283(047)\\
853.722 & 15 & 12.314(142) & 11.892(073) & 12.010(071) & 12.312(115) & .... & .... & .... & ....\\
854.764 & 30 & 12.428(119) & 11.913(064) & 12.016(058) & 12.334(098) & 12.989(078) & 13.382(035) & 12.452(068) & 12.272(092)\\
855.769 & 30 & 12.513(084) & 11.978(056) & 12.041(050) & 12.339(080) & 12.924(062) & 13.605(045) & 12.407(048) & 12.202(101)\\
856.744 & 30 & 12.672(076) & 12.045(050) & 12.080(043) & 12.342(070) & 12.875(054) & 13.584(067) & 12.377(041) & 12.143(088)\\
857.649 & 30 & .... & .... & .... & .... & 12.850(025) & .... & .... & ....\\
858.776 & 30 & 12.873(070) & 12.160(046) & 12.122(039) & 12.311(062) & 12.765(041) & 13.638(140) & 12.309(032) & 12.161(073)\\
862.723 & 30 & 13.303(066) & 12.371(048) & 12.209(039) & 12.229(058) & 12.543(056) & .... & 12.244(052) & 12.137(059)\\
863.749 & 30 & 13.409(060) & 12.442(043) & 12.240(038) & 12.221(048) & 12.506(038) & 13.505(056) & 12.207(040) & 12.069(102)\\
864.795 & 30 & 13.510(067) & 12.458(042) & 12.241(034) & 12.202(061) & 12.388(042) & .... & 12.149(045) & 12.078(100)\\
865.754 & 30 & 13.594(066) & 12.530(040) & 12.269(033) & 12.212(051) & 12.422(061) & 13.484(087) & 12.206(047) & 12.053(045)\\
866.748 & 30 & 13.695(067) & 12.580(040) & 12.322(037) & 12.207(049) & 12.328(045) & 13.414(078) & 12.163(041) & 12.039(061)\\
867.664 & 30 & 13.771(063) & 12.612(044) & 12.330(042) & 12.197(050) & 12.297(042) & 13.379(066) & 12.202(037) & 12.008(048)\\
869.736 & 30 & 13.935(065) & 12.723(044) & 12.394(034) & 12.177(049) & .... & .... & .... & ....\\
870.657 & 30 & 14.012(060) & 12.775(041) & 12.438(034) & 12.202(051) & 12.161(041) & 13.318(080) & 12.226(049) & 12.123(088)\\
871.726 & 30 & 14.087(065) & 12.826(041) & 12.481(034) & 12.215(053) & 12.094(038) & 13.278(036) & 12.219(049) & 12.147(099)\\
872.717 & 30 & 14.148(068) & 12.870(042) & 12.517(034) & 12.233(052) & 12.090(041) & 13.220(046) & 12.258(039) & 12.177(115)\\
873.732 & 30 & 14.201(291) & 12.915(121) & 12.562(042) & 12.244(112) & .... & .... & .... & ....\\
874.716 & 30 & 14.268(084) & 12.971(051) & 12.605(043) & 12.291(061) & 12.070(034) & 13.208(035) & 12.332(047) & 12.269(119)\\
875.624 & 30 & .... & .... & .... & .... & 12.038(029) & 13.239(078) & .... & ....\\
876.648 & 30 & 14.369(085) & 13.089(052) & 12.731(048) & 12.389(062) & 12.088(053) & 13.214(048) & 12.408(038) & 12.399(102)\\
877.591 & 30 & 14.401(103) & 13.160(057) & 12.779(053) & 12.444(066) & 12.123(053) & 13.270(053) & 12.448(054) & ....\\
878.630 & 30 & 14.462(125) & 13.186(064) & 12.887(058) & 12.498(071) & 12.128(044) & 13.351(087) & 12.474(059) & 12.644(170)\\
879.612 & 30 & 14.515(139) & 13.235(069) & 12.898(062) & 12.582(074) & 12.184(040) & 13.484(097) & 12.531(054) & 12.669(120)\\
880.574 & 30 & .... & .... & .... & .... & 12.212(053) & .... & 12.563(089) & ....\\
882.563 & 30 & 14.649(175) & 13.311(076) & 13.081(093) & 12.729(089) & 12.318(025) & .... & 12.759(069) & ....\\
883.596 & 30 & 14.709(258) & 13.392(109) & .... & 12.709(106) & 12.419(045) & 13.862(110) & 12.861(080) & ....\\
887.609 & 30 & 14.797(171) & 13.552(088) & 13.271(077) & 12.984(101) & 12.659(033) & 14.227(127) & 13.029(085) & 13.152(066)\\
888.593 & 30 & 14.761(218) & 13.584(101) & 13.319(085) & 13.050(104) & 12.747(052) & .... & 13.019(062) & ....\\
889.585 & 30 & 14.784(102) & 13.618(065) & 13.326(063) & 13.098(098) & 12.783(032) & 14.317(091) & .... & ....\\
890.602 & 30 & 14.790(104) & 13.629(063) & 13.381(059) & .... & 12.814(047) & 14.362(035) & 13.151(046) & 13.319(038)\\
893.627 & 30 & 14.811(114) & 13.714(070) & 13.450(066) & 13.262(100) & .... & .... & 13.309(032) & 13.549(074)\\
894.603 & 30 & 14.886(105) & 13.727(067) & 13.498(068) & 13.321(108) & 13.057(044) & 14.678(119) & 13.338(032) & ....\\
895.580 & 30 & 14.853(122) & 13.744(070) & 13.531(074) & 13.351(121) & 13.108(066) & .... & 13.425(032) & ....\\
896.560 & 30 & .... & .... & .... & .... & 13.113(025) & .... & .... & ....\\
901.563 & 30 & 14.913(129) & 13.913(078) & 13.691(075) & .... & 13.476(088) & .... & 13.569(068) & ....\\
902.645 & 90 & 14.944(116) & 13.947(149) & 13.746(128) & 13.608(121) & 13.507(054) & .... & .... & ....\\
903.720 & 90 & 14.949(108) & 13.980(070) & 13.750(071) & 13.654(136) & 13.587(087) & 15.116(035) & .... & ....\\
904.582 & 90 & 14.943(142) & 13.983(083) & 13.788(080) & 13.699(126) & 13.634(073) & .... & 13.733(045) & ....\\
905.556 & 90 & 14.960(108) & 14.006(071) & 13.836(074) & 13.700(127) & 13.638(048) & 15.222(035) & 13.791(067) & ....\\
906.539 & 90 & 14.981(299) & 14.095(140) & 13.867(138) & 13.781(221) & 13.638(031) & .... & .... & ....\\
910.542 & 90 & .... & 14.070(162) & 13.976(158) & .... & 13.876(054) & .... & .... & ....\\
911.551 & 90 & 15.025(177) & 14.142(186) & 14.028(117) & 13.958(189) & 13.889(053) & 15.598(035) & .... & ....\\
913.546 & 90 & .... & .... & .... & .... & 14.060(082) & .... & 14.103(083) & 14.290(038)\\
\enddata
\tablecomments{Julian dates refer to the median Julian dates for all the images taken in a night. Exposure times represented in the table is for $BVRJHK$ bandpasses. The exposure time for $I$ and $Y$ filter is two times longer than the exposure times listed on the table for the respective Julian dates.}
\end{deluxetable*}

\section{Results} \label{sec:results}

To analyze our supernova data, we used two different numerical models to produce fits to our lightcurves: SALT2 \citep[using SNcosmo,][]{2007A&A...466...11G} and SNooPy \citep{2011AJ....141...19B}. Each model has its own fit parameters and each model gives separate insights to the supernova and the host galaxy.
\subsection{SALT2 Modelling}
We fit our optical light curves using SALT2 \citep{2007A&A...466...11G}. Using SNcosmo \citep{2016ascl.soft11017B} we extracted values for the different fit parameters ($x_0$,$x_1$ and $c$) in the SALT2 model, where the flux of the source is given as 

\begin{multline}
F(S N, p, \lambda) = x_0 \times [M_0(p, \lambda) + x_1M_1(p, \lambda) + ...] \\ \times exp [cCL(\lambda)] 
\end{multline}

where $p$ is the rest-frame time since the date of maximum luminosity in $B$-band (the phase), and $\lambda$ the wavelength in the
rest-frame of the SN. $M_0(p, \lambda)$ is the average spectral sequence
whereas $M_k(p, \lambda)$, for $k > 0$, are additional components that describe the  variability of SNe Ia light curve. $CL(\lambda)$ represents the average color correction law. In the models from SALT2, the optical depth is expressed using a color offset with respect to the average at the date of maximum luminosity in $B$-band, $c = (B-V)_{max} - (B-V)$. This parametrization models the part of the color variation that is independent of phase, whereas the remaining color variation with phase is accounted for by the linear components. In Equation (2), $x_{0}$ is the normalization of the spectral energy distribution sequence, and $x_k$ for $k > 0$, are the intrinsic parameters of the SN (such as a stretch factor). To summarize, while ($M_k$) and $CL$ are properties of the global model, ($x_k$) and $c$ are parameters of a given supernova and hence differ from one SN to another.  

To perform accurate fits to observational data from the CTIO telescope, we registered the CTIO filters\footnote{http://www.astronomy.ohio-state.edu/ANDICAM/detectors.html} into the SNcosmo system, which uses $SDSS$ filters for its default bandpass. After registering the CTIO $BVRI$ filter band passes in the SNcosmo system, we were able to input our observed photometric magnitudes into the program to enable the modelling of the flux for the SN. 

\begin{figure*}
\plotone{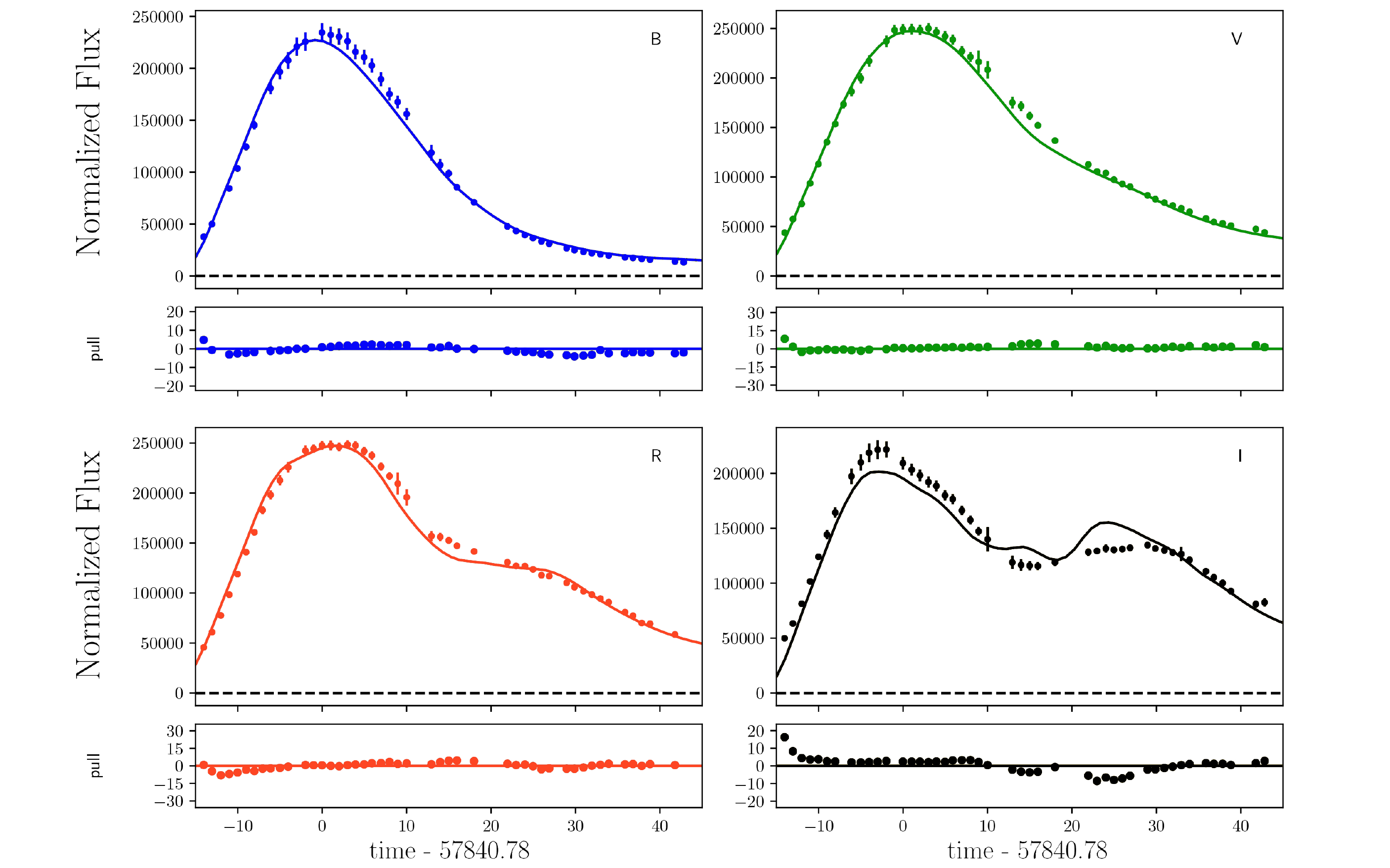}
\caption{Fits to the SN's $BVRI$ light curves using SNcosmo.\label{fig:opfit}}
\end{figure*}

Out of the 5 free parameters ($z,t_0,x_0,x_1,c$) for the SNcosmo program we set the redshift to $z \approx 0.004$ for our SN based on previous observations in the literature \citep{2004AJ....128...16K}. The fit is represented in Figure \ref{fig:opfit}. The light curve models derived the time for maximum brightness where $t_0 = 2457840.782 \pm 0.017$, corresponding to 2017 March 28, 06:46:04.8 UT. 

\subsection{SNooPy Modelling}

\begin{figure*}
\plotone{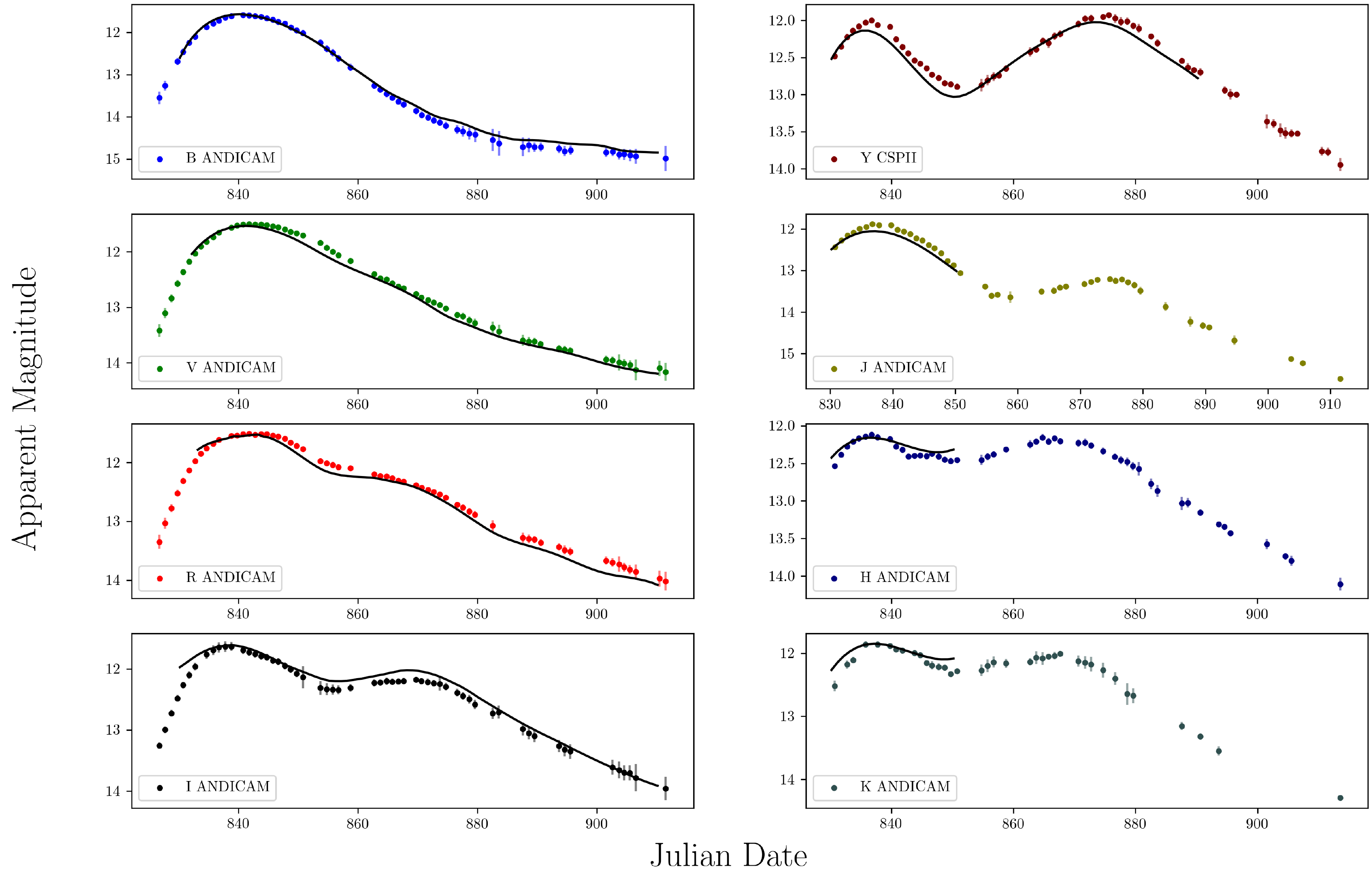}
\caption{SNooPy fits for the SN using CTIO's ANDICAM $BVRIJHK$ filters and CSPII $Y$ filter. \label{fig:irfit}}
\end{figure*}

We used the CSPII light-curve fitting code SuperNovae in Object Oriented Python \citep[SNooPy,][]{2011AJ....141...19B} to fit the light curve of SN 2017cbv in the optical and infrared. SNooPy has the prime benefit of modelling  SNe Ia in a wide range of wavelengths, as it simultaneously fits SNe Ia light curves in both the optical and near-IR. SNooPy also has a $systematics$ function, where it calculates the date of $B$-band peak $T_{max}(B)$, color excess of host galaxy $E(B-V)_{host}$, the change of $B$-band magnitude from maximum to 15 days after maximum $\Delta m_{15}(B)$, and the distance modulus. 

The model that we used in SNooPy is the $EBV_{model}$. The light curve model is a variation of the model given by \citet{2006ApJ...647..501P}:

\begin{multline*}
m_X(t-t_{max}) = T_Y((t'-t_{max})/(1+z),\Delta m_{15}) + M_Y(\Delta m_{15}) \\
+ \mu + R_XE(B-V)_{gal} + R_YE(B-V)_{host} + \\ K_{X,Y} (z,(t'-t_{max})/(1+z),E(B-V)_{host},E(B-V)_{gal})
\end{multline*}

where $m_X$ is the observed magnitude in band $X$, $t_{max}$ is the time of $B$ maximum, $\Delta m_{15}$ is the decline rate parameter \citep{1993ApJ...413L.105P}, $M_Y$ is the absolute magnitude in filter $Y$ in the rest-frame of supernova, $E(B-V)_{gal}$ and $E(B-V)_{host}$ are reddening due to galactic foreground and host galaxy, respectively, $R_X$ and $R_Y$ are the total-to-selective absorptions for filters $X$ and $Y$, respectively, $K_{XY}$ is the cross-band $K$-correction from rest-frame $X$ to observed filter $Y$. SNooPy uses the Schlegel maps for the handling of galactic extinction \citep{2011AJ....141...19B}. 

For the fit, we registered the CTIO $BVRJHK$ filters to SNooPy which uses CSPII filters as its default bandpass. As there was no transmission data for ANDICAM $Y$ bandpass, we used the CSPII $Y$ filter for the fit for our $Y$ bandpass. The resulting fit with SNooPy presented in Figure 5.

With CTIO's filter transmission data and our $BVR$ $YJHK$ light curves, SNooPy's $systematics$ function yields the following results. $T_{max}(B)$ is on 2017 March 28, 11:09:36.0 UT (JD = 2457840.965 $\pm$ 0.088 with systematic model error of 0.340). The color excess of host galaxy is determined to be $-0.090 \pm 0.009$ with a systematic model error of $\pm$ 0.060. The $\Delta m_{15}$ is 0.877 $\pm$ 0.010 with a systematic model error of $\pm$ 0.060. The distance modulus is 30.499 $\pm$ 0.008 with a systematic model error of $\pm$ 0.309.

\section{Discussion} \label{sec:Discussion}
\subsection{Optical Light Curves}
A typical Type Ia supernova has a $B$-band decline rate $\Delta m_{15}$(B) = 1.1 mag \citep{1999AJ....118.1766P}, and this parameter varies roughly from 0.75 to 1.94 \citep{2003AJ....125..166K}. Compared to the $\Delta m_{15}$(B) variation, SN 2017cbv has a broad decline of $\Delta m_{15}$(B) = 0.877 $\pm$ 0.070 mag (including systematic model error) according to the SNooPy model. 

As expected of optical light curves of SNe Ia, the $B$-band decline rate of the SN is the most rapid, followed by $V$-band, $R$-band, and $I$-band respectively. We also see the expected secondary bump in the $I$ optical band and a “shoulder” in the $R$ band at JD $\approx 2457869.70$.

The relatively faster $B$-band decline rate for SNe Ia is caused by the increasing absorption by Fe $\textsc{ii}$ and Co \textsc{ii} lines as the ejecta cools, which blocks transmissivity of bluer wavelength bands and simultaneously improve transparency at a longer wavelength post maximum light of SNe Ia \citep{2006ApJ...649..939K}. 

\subsection{Near-Infrared Light Curves}
Our near-IR $YJHK$ light curves shows the expected secondary maximum for near-IR wavelengths. The secondary maximum is noteworthy for the $Y$ and $H$ bands, where the secondary peaks are brighter than the primary peaks, a phenomenon not uncommon in SNe Ia's near-IR light curves (Phillips et al. 2013).

The mechanics of the secondary maximum in the near-IR is similar to that of the $I$ and $R$ bump in Kasen's models \citep{2006ApJ...649..939K}, where the ejecta cools and becomes transparent in the near-IR, allowing photons in the near-IR wavelengths to escape the ejecta. Future work can shed light on the astrophysics of the secondary bump by comparing the shape of the SN's high cadence light curves with other SNe Ia.

\begin{figure*}
\plotone{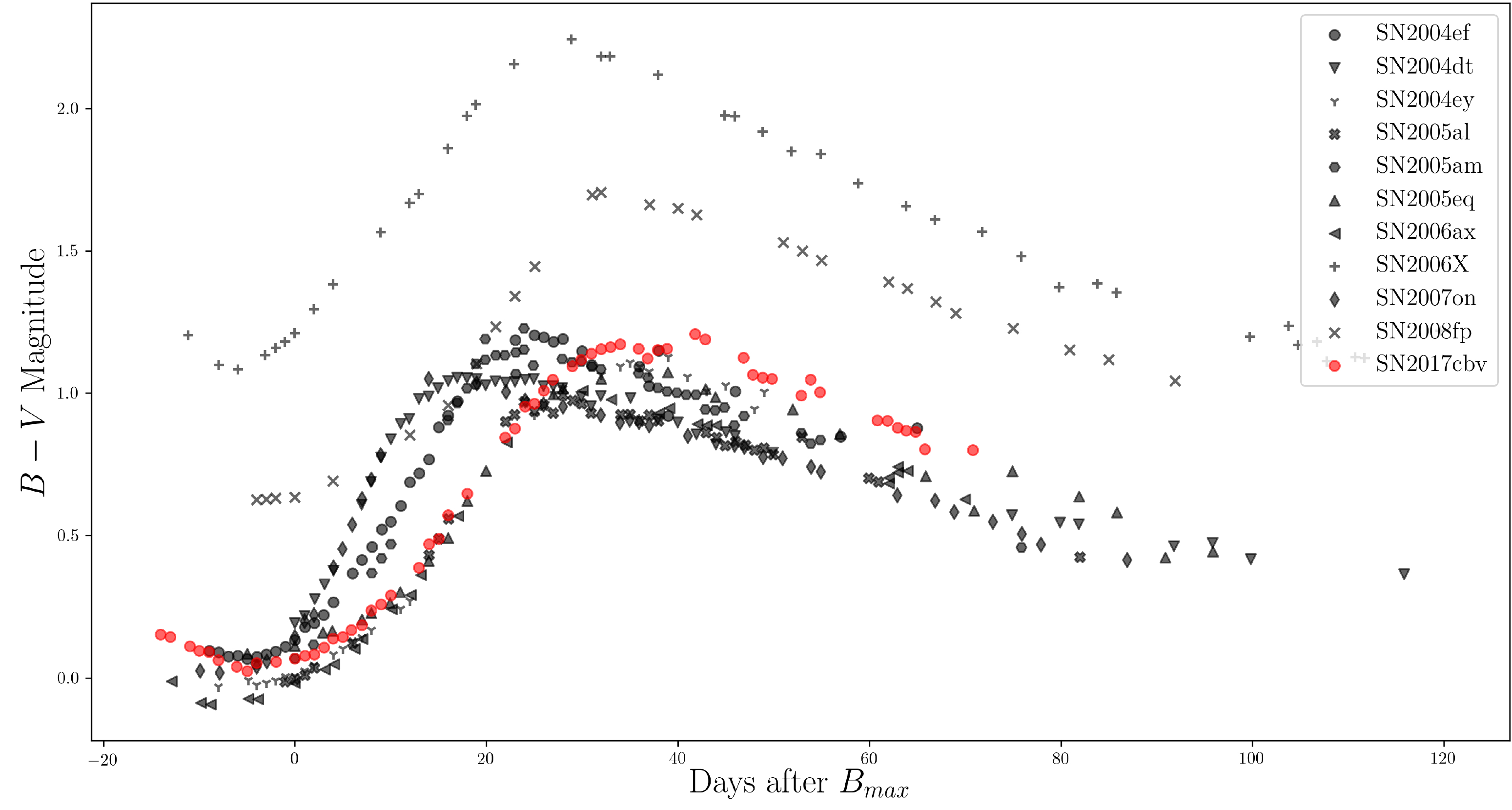}
\caption{$B-V$ color curve of SN 2017cbv in comparison with photometric data of  low-$z$ SNe Ia.
\label{fig:BV}}
\end{figure*}
\subsection{Reddening and Extinction}

From 30 to 90 days after $V$-band maximum, the color evolution for unreddened $B-V$ colors of SNe Ia is thought to be uniform \citep{1995} in what is called the Lira law regime. The spectra of the SN during Lira law regime is uniform throughout the late epoch as the Fe-Ni-Co core dominates the spectrum that is rapidly evolving into a supernebula phase \citep{2006ApJ...649..939K}. Knowing the intrinsic $B-V$ colors of a Type Ia supernova allows us to calculate the extinction of the SN, in the form of the ratio of total-to-selective dust absorption, $R_V$. Figure \ref{fig:BV} shows the $B-V$ color curve of SN 2017cbv. 

As SNe Ia have varying $B-V$ colors throughout the evolution of their light curves, there are several views as to which part of the $B-V$ curve provides us with the most meaningful $B-V$ value to use for calculating color excess. \citet{2007ApJ...659..122J}, through an analysis of a large sample of SNe Ia, found that the intrinsic colors of SNe Ia are identical and well-described by a Gaussian distribution at 35 days post $B$-band maximum. We take Jha's intrinsic $B-V$ of 1.054 $\pm$ 0.049 mag at t\textsubscript{35} \citep{2007ApJ...659..122J}. Since our observed $B-V$ at t\textsubscript{35} is 1.164 $\pm$ 0.08 mag, the color excess $E(B-V)$ of the SN at t\textsubscript{35} is 0.11 $\pm$ 0.129. With the Galactic reddening of 0.145 $\pm$ 0.001 in the coordinates of the SN \citep{2011ApJ...737..103S}, the host galaxy reddening is negligible. The findings are consistent with SNooPy's $systematics$ function, where the function also derived zero host galaxy reddening using the Schlegel maps. Since extinction is negligible ($A_V \approx 0$), it is not possible to determine the total-to-selective ratio $R\textsubscript{V}$ of the host galaxy NGC5643. 

\subsection{Distance}
SNooPy's $systematics$ function provides a distance modulus to the SN using the calibrated absolute magnitudes of the SN using Prieto et al.'s method \citep{2006ApJ...647..501P} as stated above. Prieto's method uses the concordance model and assumes $H_0$ = 72 km s\textsuperscript{-1} Mpc\textsuperscript{-1} \citep{2006ApJ...647..501P}.  The averaged distance modulus $(m-M)$\textsubscript{BVRI} of the SN based on SNooPy's $systematics$ function is 30.499 $\pm$ 0.008 with a systematic model error of $\pm$ 0.309, corresponding to a distance of $12.583_{-1.709}^{+1.977}$ Mpc. The distance modulus is in agreement with the Galactocentric GSR distance of $(m-M)$ = 30.83 $\pm$ 0.15 and the average redshift-independent distance $(m-M)$ = 30.25 $\pm$ 0.44 to the host galaxy NGC5643. The information on the distances is obtained from the NASA/IPACExtragalactic Database. The redshift-independent distance and the Galactocentric GSR distance for NGC5643 disagree with each other, but interestingly, our results fall within the range of both distances.

\section{Conclusion}
We have presented optical and near-IR photometry of SN 2017cbv. We found that the SN is a broad decliner with a $\Delta m_{15}(B) = 0.877 \pm 0.070$ mag. The SN also exhibit the typical SNe Ia light curves, with the rapid decline of $B$ and $V$ bands, and the slight secondary bump in the $R$ and $I$ in the optical light curves. We also observe the secondary maximums for the $Y$ and $H$ bands that are brighter than their primary maximums, and the characteristic secondary bump in the $J$ and $K$ bands. From the $\Delta m_{15}$, we estimated $M_V$ \citep{2006ApJ...647..501P, 1993ApJ...413L.105P} and calculated $M_V = -19.380 \pm 0.027$ for the SN. From SALT2, we derived the time of maximum brightness corresponded to be $t_0(B) = 2457840.782 \pm 0.017$, which is in agreement with SNooPy's figure of $T_{max}(B) = 2457840.965 \pm 0.088 \pm 0.340$.

With the Schlegel and Schlafly maps, we also find that SN 2017cbv was unreddenned by host galaxy dust, where $E(B-V)_{host} \approx 0$. Since host galaxy extinction is negligible, the total-to-selective ratio $R_V$ cannot be determined. We also derived a distance modulus of the object where $(m-M) =  30.499 \pm 0.008 \pm 0.309$, corresponding to a distance of $12.583_{-1.709}^{+1.977} $  Mpc. The distance of the SN is consistent with both redshift-dependent and redshift-independent derivations of the distance to its host galaxy. 

Considering the proximity of the SN, our distance determination to the SN has important implications for cosmologists working on the precision of the Hubble constant. Notably, our distance determination could assist the Carnegie-Chicago Hubble Program \citep{2016ApJ...832..210B}, which relies on RR Lyrae zero-point, the TRGB method, and local SNe Ia that is not in the Hubble flow where $z < 0.01$. A comparison of the SN's $B-V$ color curves with CSPII photometric data of ten other low-$z$ SNe Ia \citep{2011AJ....142..156S} in Figure \ref{fig:BV} shows that the SN conforms to the usual color evolution of other SNe Ia, which makes it ideal for further studies.

\acknowledgments

We like to thank Mark Phillips for reading through and providing critical comments for the draft of our paper and Rohan Naidu for his help in developing new Python routines for analysis of the optical photometry of this project. We also like to thank Rachael Beaton and Ben Shappee for discussions on the supernova and Chris Burns for his help with configuring SNooPy for our supernova modelling. 

This work is supported and funded by Yale-NUS College and JY Pillay Foundation. B.E.P. would like to acknowledge the support from Yale-NUS College, and from funding from NSF grant (AST-1440341) and an NSF PIRE Grant 1545949.

%

\vspace{5mm}
\facilities{CTIO 1.3m, LCRO 0.3m}


\software{Astropy \citep{2013A&A...558A..33A},  
          Cyanogen Imaging MaxIm DL v6 (\url{http://diffractionlimited.com/product/maxim-dl/}), SNCosmo \citep{2016ascl.soft11017B}, SNooPy \citep{2011AJ....141...19B}
          }




\newpage
\begin{thebibliography}{}

\bibitem[Astropy Collaboration et al.(2013)]{2013A&A...558A..33A} Astropy Collaboration, Robitaille, T.~P., Tollerud, E.~J., et al.\ 2013, \aap, 558, A33 
\bibitem[Antognini, et al.(2014)]{2014MNRAS.439.1079A} Antognini, J.~M., Shappee, B.~J., Thompson, T.~A., et al.\ 2014, \mnras, 439, 1079.
\bibitem[Barbary et al.(2016)]{2016ascl.soft11017B} Barbary, K., Barclay, T., Biswas, R., et al.\ 2016, Astrophysics Source Code Library, ascl:1611.017 
\bibitem[Beaton et al.(2016)]{2016ApJ...832..210B} Beaton, R.~L., Freedman, W.~L., Madore, B.~F., et al.\ 2016, \apj, 832, 210 
\bibitem[Burns et al.(2011)]{2011AJ....141...19B} Burns, C.~R., Stritzinger, M., Phillips, M.~M., et al.\ 2011, \aj, 141, 19 
\bibitem[Bradley et al.(2016)]{2016ascl.soft09011B} Bradley, L., Sipocz, B., Robitaille, T., et al.\ 2016, Astrophysics Source Code Library, ascl:1609.011
\bibitem[Cardelli et al.(1989)]{1989ApJ...345..245C} Cardelli, J.~A., Clayton, G.~C., \& Mathis, J.~S.\ 1989, \apj, 345, 245 
\bibitem[Drake, et al.(2009)]{2009ApJ...696..870D} Drake, A.~J., Djorgovski, S.~G., Mahabal, A., et al.\ 2009, \apj, 696, 870.
\bibitem[Drake, et al.(2017)]{2017MNRAS.469.3688D} Drake, A.~J., Djorgovski, S.~G., Catelan, M., et al.\ 2017, \mnras, 469, 3688.
\bibitem[Elias et al.(1985)]{1985ApJ...296..379E} Elias, J.~H., Matthews, K., Neugebauer, G., \& Persson, S.~E.\ 1985, \apj, 296, 379  
\bibitem[Germany et al.(2004)]{2004A&A...415..863G} Germany, L.~M., Reiss, D.~J., Schmidt, B.~P., Stubbs, C.~W., \& Suntzeff, N.~B.\ 2004, \aap, 415, 863 
\bibitem[Guy et al.(2007)]{2007A&A...466...11G} Guy, J., Astier, P., Baumont, S., et al.\ 2007, \aap, 466, 11 
\bibitem[Gilfanov \& Bogd{\'a}n(2010)]{2010Natur.463..924G} Gilfanov, M., \& Bogd{\'a}n, {\'A}.\ 2010, \nat, 463, 924 
\bibitem[Hamuy et al.(1996)]{1996AJ....112.2391H} Hamuy, M., Phillips, M.~M., Suntzeff, N.~B., et al.\ 1996, \aj, 112, 2391 
\bibitem[Hatt et al.(2017)]{2017ApJ...845..146H} Hatt, D., Beaton, R.~L., Freedman, W.~L., et al.\ 2017, \apj, 845, 146 
\bibitem[Hillebrandt \& Niemeyer(2000)]{2000ARA&A..38..191H} Hillebrandt, W., \& Niemeyer, J.~C.\ 2000, \araa, 38, 191 
\bibitem[Iben \& Tutukov(1984)]{1984ApJS...54..335I} Iben, I., Jr., \& Tutukov, A.~V.\ 1984, \apjs, 54, 335 
\bibitem[Jang et al.(2017)]{2017arXiv170310616J} Jang, I.~S., Hatt, D., Beaton, R.~L., et al.\ 2017, arXiv:1703.10616 
\bibitem[Jha et al.(1999)]{1999ApJS..125...73J} Jha, S., Garnavich, P.~M., Kirshner, R.~P., et al.\ 1999, \apjs, 125, 73 
\bibitem[Jha et al.(2007)]{2007ApJ...659..122J} Jha, S., Riess, A.~G., \& Kirshner, R.~P.\ 2007, \apj, 659, 122 
\bibitem[Kasen(2006)]{2006ApJ...649..939K} Kasen, D.\ 2006, \apj, 649, 939 
\bibitem[Koribalski et al.(2004)]{2004AJ....128...16K} Koribalski, B.~S., Staveley-Smith, L., Kilborn, V.~A., et al.\ 2004, \aj, 128, 16
\bibitem[Krisciunas et al.(2000)]{2000ApJ...539..658K} Krisciunas, K., Hastings, N.~C., Loomis, K., et al.\ 2000, \apj, 539, 658 
\bibitem[Krisciunas et al.(2003)]{2003AJ....125..166K} Krisciunas, K., Suntzeff, N.~B., Candia, P., et al.\ 2003, \aj, 125, 166 
\bibitem[Krisciunas et al.(2007)]{2007AJ....133...58K} Krisciunas, K., Garnavich, P.~M., Stanishev, V., et al.\ 2007, \aj, 133, 58 
\bibitem[Krisciunas et al.(2017)]{2017AJ....154..211K} Krisciunas, K., Contreras, C., Burns, C.~R., et al.\ 2017, \aj, 154, 211 
\bibitem[Landolt(1992)]{1992AJ....104..340L} Landolt, A.~U.\ 1992, \aj, 104, 340 
\bibitem[Lang et al.(2010)]{2010AJ....139.1782L} Lang, D., Hogg, D.~W., Mierle, K., Blanton, M., \& Roweis, S.\ 2010, \aj, 139, 1782 
\bibitem[Liu et al.(2010)]{2010A&A...523A...3L} Liu, W.-M., Chen, W.-C., Wang, B., \& Han, Z.~W.\ 2010, \aap, 523, A3 
\bibitem[Lira (1995)]{1995} Lira, ~P.\ 1995, Masters Thesis, Universidad de Chile
\bibitem[Mandel et al.(2009)]{2009ApJ...704..629M} Mandel, K.~S., Wood-Vasey, W.~M., Friedman, A.~S., \& Kirshner, R.~P.\ 2009, \apj, 704, 629 
\bibitem[Moll et al.(2014)]{2014ApJ...785..105M} Moll, R., Raskin, C., Kasen, D., \& Woosley, S.~E.\ 2014, \apj, 785, 105 
\bibitem[Munari \& Renzini(1992)]{1992ApJ...397L..87M} Munari, U., \& Renzini, A.\ 1992, \apjl, 397, L87
\bibitem[Nomoto(1982)]{1982ApJ...253..798N} Nomoto, K.\ 1982, \apj, 253, 798 
\bibitem[Nomoto et al.(1984)]{1984ApJ...286..644N} Nomoto, K., Thielemann, F.-K., \& Yokoi, K.\ 1984, \apj, 286, 644 
\bibitem[Pejcha, et al.(2013)]{2013MNRAS.435..943P} Pejcha, O., Antognini, J.~M., Shappee, B.~J., et al.\ 2013, \mnras, 435, 943.
\bibitem[Persson et al.(1998)]{1998AJ....116.2475P} Persson, S.~E., Murphy, D.~C., Krzeminski, W., Roth, M., \& Rieke, M.~J.\ 1998, \aj, 116, 2475 
\bibitem[Podsiadlowski et al.(2008)]{2008NewAR..52..381P} Podsiadlowski, P., Mazzali, P., Lesaffre, P., Han, Z., \& F{\"o}rster, F.\ 2008, \nar, 52, 381 
\bibitem[Phillips(1993)]{1993ApJ...413L.105P} Phillips, M.~M.\ 1993, \apjl, 413, L105 
\bibitem[Phillips et al.(1999)]{1999AJ....118.1766P} Phillips, M.~M., Lira, P., Suntzeff, N.~B., et al.\ 1999, \aj, 118, 1766 
\bibitem[Phillips(2005)]{2005ASPC..342..211P} Phillips, M.~M.\ 2005, 1604-2004: Supernovae as Cosmological Lighthouses, 342, 211 
\bibitem[Phillips et al.(2013)]{2013ApJ...779...38P} Phillips, M.~M., Simon, J.~D., Morrell, N., et al.\ 2013, \apj, 779, 38 
\bibitem[Prieto et al.(2006)]{2006ApJ...647..501P} Prieto, J.~L., Rest, A., \& Suntzeff, N.~B.\ 2006, \apj, 647, 501 
\bibitem[Perlmutter \& Riess(1999)]{1999AIPC..478..129P} Perlmutter, S., \& Riess, A.\ 1999, COSMO-98, 478, 129 
\bibitem[Riess et al.(1996)]{1996ApJ...473...88R} Riess, A.~G., Press, W.~H., \& Kirshner, R.~P.\ 1996, \apj, 473, 88 
\bibitem[Riess et al.(1998)]{1998AJ....116.1009R} Riess, A.~G., Filippenko, A.~V., Challis, P., et al.\ 1998, \aj, 116, 1009 
\bibitem[R{\"o}pke et al.(2012)]{2012ApJ...750L..19R} R{\"o}pke, F.~K., Kromer, M., Seitenzahl, I.~R., et al.\ 2012, \apjl, 750, L19 
\bibitem[Schlafly \& Finkbeiner(2011)]{2011ApJ...737..103S} Schlafly, E.~F., \& Finkbeiner, D.~P.\ 2011, \apj, 737, 103
\bibitem[Schlegel et al.(1998)]{1998ApJ...500..525S} Schlegel, D.~J., Finkbeiner, D.~P., \& Davis, M.\ 1998, \apj, 500, 525 
\bibitem[Shappee \& Thompson(2013)]{2013ApJ...766...64S} Shappee, B.~J. \& Thompson, T.~A.\ 2013, \apj, 766, 64.
\bibitem[Sternberg et al.(2014)]{2014MNRAS.443.1849S} Sternberg, A., Gal-Yam, A., Simon, J.~D., et al.\ 2014, \mnras, 443, 1849 
\bibitem[Stritzinger et al.(2011)]{2011AJ....142..156S} Stritzinger, M.~D., Phillips, M.~M., Boldt, L.~N., et al.\ 2011, \aj, 142, 156 
\bibitem[van den Heuvel et al.(1992)]{1992A&A...262...97V} van den Heuvel, E.~P.~J., Bhattacharya, D., Nomoto, K., \& Rappaport, S.~A.\ 1992, \aap, 262, 97 
\bibitem[VanderPlas, et al.(2014)]{2014ascl.soft07018V} VanderPlas, J., Fouesneau, M. \& Taylor, J.\ 2014, Astrophysics Source Code Library , ascl:1407.018.
\bibitem[Webbink(1984)]{1984ApJ...277..355W} Webbink, R.~F.\ 1984, \apj, 277, 355 
\bibitem[Wood-Vasey et al.(2008)]{2008ApJ...689..377W} Wood-Vasey, W.~M., Friedman, A.~S., Bloom, J.~S., et al.\ 2008, \apj, 689, 377-390 
\bibitem[Woosley \& Weaver(1986)]{1986LNP...255...91W} Woosley, S.~E., \& Weaver, T.~A.\ 1986, IAU Colloq.~89: Radiation Hydrodynamics in Stars and Compact Objects, 255, 91 
\bibitem[Yoon \& Langer(2003)]{2003A&A...412L..53Y} Yoon, S.-C., \& Langer, N.\ 2003, \aap, 412, L53 












\end{thebibliography}
\end{document}